\documentclass[epjc,preprint]{revtex4}
\usepackage{amsfonts}
\usepackage{amsmath}
\usepackage{amssymb}
\usepackage{graphicx}
\usepackage{float}
\usepackage{caption}
\usepackage{subcaption}

\begin{document}

\title{The influence of lunar tidal potential on clock frequencies at
different positions on Earth}
\author{Hongbin Zhang}
\author{Yanyue Gao}
\author{Baocheng Zhang}
\email{zhangbaocheng@cug.edu.cn}
\affiliation{School of Mathematics and Physics, China University of Geosciences, Wuhan
430074, China}

\begin{abstract}
With the advancements in clock timing technology, increasingly smaller time
differences can be distinguished. Therefore, it is critical to investigate
the fractional frequency shift of clocks at different locations on Earth. In
this paper, we study it systematically under the influence of a subtle lunar
tidal potential based on a new method. Our calculations in the geocentric
Fermi frame show that when two clocks are located at the same latitude, the
longitude difference changes the fractional frequency shift between them. A
similar phenomenon occurs when there is a difference in latitude between two
clocks on the ground at the same longitude. Interestingly, when the Moon's
longitude changes, the phase and amplitude of the lunar tidal fractional
frequency shift between two clocks with the same longitude difference will
change, while the change in the Moon's latitude only affects the amplitude
of the fractional frequency shift of these two clocks. Our results provide
useful information for the calibration and synchronization of clocks on
Earth. \newline

\end{abstract}

\maketitle

\section{Introduction}

After decades of development, clock performance has reached unprecedented
precision, with instabilities and uncertainties reaching $10^{-18}$ \cite%
{bloom2014optical,nicholson2015systematic,huntemann2016single,nakamura2020coherent,boulder2021frequency}
or better \cite{marshall2025high} in fractional frequency. This improvement
in performance has led to a wide range of applications for high-precision
clocks, such as detecting gravitational redshift \cite%
{delva2018gravitational,herrmann2018test,takamoto2020test,shen2023testing},
detecting dark matter \cite{derevianko2014hunting,bhardwaj2017situ}, and
verifying the equivalence principle \cite%
{wolf2006cold,pihan2017lorentz,sanner2019optical,qin2021test}. The accuracy
of the clock is crucial for these experiments. Another important application
that places high demands on clock performance is communication and
navigation, which rely on clock networks that are synchronized to within
tens of nanoseconds by making clock comparisons.

The International Astronomical Union (IAU) resolution provides a fully
relativistic framework for transformations between coordinates and
gravitational potentials, and uses post-Newtonian potentials to parameterize
the potential coefficients to construct local reference systems for all
celestial bodies in the Solar System \cite{soffel2003iau,kaplan2006iau}. On
this basis, an equation for comparing the rates of clocks on the lunar
surface relative to geocentric time can be derived \cite{kopeikin2024lunar}.
The concept of a generalized Fermi frame was constructed to describe the
relativistic effects of distant celestial bodies like the Moon on the motion
of the Earth's satellites \cite{fermi1922phenomena,ashby1986relativistic}.
The generalized Fermi frame is suitable for comparing clock rates at the
Moon and the Earth-Moon Lagrange point relative to clocks on Earth using a
local free-fall frame \cite{ashby2024relativistic}. Both methods are
suitable for describing the gravitational effects on clocks near the Earth.
Here, we use the latter method to analyze the frequency shift of clocks on
Earth in the first post-Newtonian approximation.

The time measured by a clock at any given location is the proper time. There
are three main factors that can cause the clock frequency shift: the motion
of the clock, the position of the clock in the main gravitational potential,
and the tidal potential. Blanchet et al. took into account the motion effect
of the clock and the gravitational potential of the Earth in \cite%
{blanchet2001relativistic}. According to the equivalence principle in the
local coordinate system, the influence of external matter should be given by
the tidal potential, and the intrinsic effect of external matter is the
inhomogeneous gravitational field \cite{soffel2003iau,poisson2014gravity}.
The exact expression for the tidal potential can be found in \cite%
{murray1999solar}. Full post-Newtonian expressions for the tidal potential
can be found in Damour et al. \cite{damour1992general}. Kozai \cite{Kozai}
gave the tidal potential of the Moon as felt by clocks on Earth's satellites.

Gravitational and kinematic corrections in the frequency shift on the order
of $c^{-2}$ have already been determined by Vessot et al. \cite%
{vessot1980test}. Linet et al. \cite{linet2002time} further developed the
relativistic frequency shift in the field of an axisymmetric rotating body.
Numerical estimates of the contribution of the tidal potential to clock
frequency shift were provided by Wolf et al. \cite{wolf1995relativistic}.
Qin et al. \cite{qin2019relativistic,qin2020tidal,qin2023tidal} reported the
fractional frequency shift between two clocks at different latitudes on the
Earth's surface due to the tidal potentials of the Moon and the Sun. Zhang
et al. \cite{zhang2025frequency} considered the fractional frequency shift
between clocks on the lunar surface and clocks on the Earth's surface caused
by both tidal and non-tidal potential. The measurement of tidal effects in
clock comparison is important for both general relativity and the
equivalence principle. Compared to the Earth's gravitational field, the
frequency shifts caused by tides are much smaller. Tidal potential is
usually neglected in the local Fermi frame \cite{ashby2024relativistic}.
With the improvements in clock performance, it is now possible to measure
the frequency shifts caused by lunar tides. In the present work, in contrast
to the analysis by Qin et al. \cite{qin2020tidal} and Zhang et al. \cite%
{zhang2025frequency} on the fractional frequency shift of clocks in the
geocentric coordinate reference system (GCRS), we hope to obtain the
fractional frequency shift of clocks caused by both tidal potential and
non-tidal potential in the geocentric Fermi frame.

This paper is organized as follows. In the second section, we present the
proper time of clocks in the geocentric Fermi frame that includes the
Earth's gravitational potential, lunar tidal potential, and the tidal
potential of the Sun. In the third section, we present the fractional
frequency shift for two clocks at different locations on the Earth. The
different locations of the two clocks and the motion of the Moon affect not
only the magnitude of the lunar tidal fractional frequency shift but also
its phase. Finally, we present our conclusions in the fourth section.

\section{Clocks in the geocentric Fermi frame}

The equivalence principle of general relativity states that all coordinate
systems are mathematically equivalent, allowing researchers to select any
coordinate system for analyzing experimental observables \cite%
{turyshev2013general}. Choosing an appropriate coordinate system simplifies
physical models. Different reference systems help to address specific
practical problems. The barycentric coordinate reference system (BCRS), a
global frame centered at the Solar System's mass center, is crucial for
studying the dynamical effects of celestial bodies within the Solar System.
In contrast, the GCRS, a local frame centered at Earth's mass center, is
appropriate for studying gravitational and tidal effects near Earth. While
the Solar System BCRS provides a global perspective for large-scale
dynamics, the GCRS is more suitable for processes localized to in the
vicinity of Earth. For some specific observers, these two reference frames
are effectively equivalent \cite{ries1988effect}.

Here, we use geocentric Fermi coordinates to the order $c^{-2}$ \cite%
{ashby2024relativistic} similar to the GCRS:
\begin{equation}
-ds^{2}=-\left( 1+\frac{2\Phi _{e}}{c^{2}}+\frac{2\Phi _{tm}}{c^{2}}+\frac{%
2\Phi _{ts}}{c^{2}}\right) (dx^{0})^{2}+\left( 1-\frac{2\Phi _{e}}{c^{2}}-%
\frac{2\Phi _{tm}}{c^{2}}-\frac{2\Phi _{ts}}{c^{2}}\right)
(dx^{2}+dy^{2}+dz^{2}),  \label{gF}
\end{equation}%
where the lowercase letter $dx^{0}$ is the time coordinate displacement, and
$dx,dy,dz$ are the space coordinate displacements in the geocentric Fermi
frame. $\Phi _{e}$ is the gravitational potential of the Earth, $\Phi _{tm}$
is the tidal potential of the Moon, and $\Phi _{ts}$ the tidal potential of
the Sun. When a clock is fixed on Earth, its proper time is related to the
proper time of a clock infinitely far away from the mass center of the
Earth, or relative coordinate time, as follows
\begin{equation}
-c^{2}d\tau ^{2}=-ds^{2}=-\left( 1+\frac{2\Phi _{e}}{c^{2}}|_{R}+\frac{2\Phi
_{tm}}{c^{2}}|_{R}+\frac{2\Phi _{ts}}{c^{2}}|_{R}-\frac{v^{2}}{c^{2}}\right)
c^{2}dt^{2},  \label{pc}
\end{equation}%
where the notation $|_{R}$ represents the potentials evaluated at the
location of the clock, and $v^{2}=\left( dx^{2}+dy^{2}+dz^{2}\right) /dt^{2}$
that $v$ is the clock speed in the geocentric Fermi frame. Equation (\ref{pc}%
) is obtained by retaining the terms of order $c^{-2}$ from Eq. (\ref{gF}),
and can be further rewritten and expanded as
\begin{equation}
cd\tau =\sqrt{1+K}cdt=\left( 1+\frac{K}{2}-\frac{K^{2}}{8}+\cdots \right)
cdt,  \label{tc}
\end{equation}%
where $K=2\Phi _{e}/c^{2}|_{R}+2\Phi _{tm}/c^{2}|_{R}+2\Phi
_{ts}/c^{2}|_{R}-v^{2}/c^{2}$. Note that Eq. (\ref{tc}) is only accurate up
to the order $c^{-2}$, as in Eq. (\ref{pc}). We then gain the proper time of
the clock in the geocentric Fermi frame that includes the Earth's
gravitational potential, the Moon's tidal potential and the Sun's tidal
potential,
\begin{equation}
d\tau =\left( 1+\frac{\Phi _{e}}{c^{2}}|_{R}+\frac{\Phi _{tm}}{c^{2}}|_{R}+%
\frac{\Phi _{ts}}{c^{2}}|_{R}-\frac{v^{2}}{2c^{2}}\right) dt.  \label{pc2}
\end{equation}

In this paper, we will focus on the influence of lunar tidal potential on
the frequency change of clocks. The tidal potential of the Moon $\Phi _{tm}$
is usually obtained by expanding the gravitational potential of the Moon $%
\Phi _{m}=-\left( GM_{m}\right) /\triangle $, where
\begin{equation}
\triangle =r_{me}\left[ 1-2\left( \frac{r}{r_{me}}\right) \cos \psi +\left(
\frac{r}{r_{me}}\right) ^{2}\right] ^{\frac{1}{2}},
\end{equation}%
is the distance between observers on the Earth and the Moon, and the angle $%
\psi $ is the angle between the position vector $\mathbf{r}$ and the
position vector $\mathbf{r_{me}}$. Because $r/r_{me}\leq 1$, the expanded
form of the Moon's gravitational potential can be expressed as
\begin{eqnarray}
\Phi _{m} &=&-\frac{GM_{m}}{r_{me}}[1+\left( \frac{r}{r_{me}}\right) \cos
\psi +\left( \frac{r}{r_{me}}\right) ^{2}\frac{1}{2}\left( 3\cos ^{2}\psi
-1\right)   \notag \\
&&+\left( \frac{r}{r_{me}}\right) ^{3}\frac{1}{2}\left( 5\cos ^{3}\psi
-3\cos \psi \right) +\cdots ].  \label{gp}
\end{eqnarray}%
The first term in Eq. (\ref{gp}), $-GM_{m}/r_{me}$, is a constant term. The
second term, $-\left( GM_{m}r\cos \psi \right) /r_{me}^{2}$, is the
Newtonian gravitational potential of the Moon. The tidal potential of the
Moon is \cite{murray1999solar}
\begin{equation}
\Phi _{tm}=-\frac{GM_{m}r^{2}}{r_{me}^{3}}\frac{1}{2}\left( 3\cos ^{2}\psi
-1\right) ,  \label{tm}
\end{equation}%
where the higher-order terms in the expansion have been neglected, because
they decrease as the order of expansion increases. For example, the
magnitude of a third-order term is approximately $1\times 10^{-19}$, which
is much smaller than that of a second-order term, $1\times 10^{-17}$. All
calculations and analyses in the next section will remain at the level of $%
10^{-18}$ according to the measurement precision of the clock at present.

Note that the influence on the clock frequency from the Earth's
gravitational potential and the Sun's tidal potential can reach the level of
$10^{-17}$. Here, we present a brief analysis of these two terms. The
Earth's gravitational potential can be expressed as \cite%
{JPL-EarthGrav:2021,Montenbruck-Gill:2012}
\begin{equation}
\Phi _{e}=\frac{GM_{e}}{r}\sum_{l=0}^{\infty }\left( \frac{a_{e}}{r}\right)
^{l}\sum_{m=0}^{l}\bar{P}_{l,m}(\sin \phi )\left[ \bar{C}_{l,m}\cos \left(
m\lambda \right) +\bar{S}_{l,m}\sin \left( m\lambda \right) \right] ,
\label{pe}
\end{equation}%
where $GM_{e}$ is the product of Earth's mass and the Newtonian
gravitational constant, $a_{e}$ is Earth's equatorial radius, $r,\phi
,\lambda $ are the distance from a point on the Earth's surface to the
Earth's center of mass, the latitude of the point on the Earth's surface,
and the longitude, respectively. $\bar{C}_{l,m}$ and $\bar{S}_{l,m}$ are
normalized spherical harmonic coefficients of degree $l$ and order $m$, and $%
\bar{P}_{l,m}$ are the normalized associated Legendre functions of degree $l$
and order $m$. Their detailed expressions can be found in Refs. \cite%
{JPL-EarthGrav:2021,Montenbruck-Gill:2012}. Similar to lunar tidal
potential, the solar tidal potential can be expressed as \cite%
{ashby2024relativistic}
\begin{equation}
\Phi _{ts}=-\frac{GM_{s}r^{2}}{r_{se}^{3}}\frac{1}{2}\left( 3\cos ^{2}\psi
_{s}-1\right) ,  \label{ts}
\end{equation}%
where $M_{s}$ is the solar mass, $r_{se}$ is the distance from the Earth to
the Sun, and the angle $\psi _{s}$ is the angle between the position vector $%
\mathbf{r}$ and the position vector $\mathbf{r_{se}}$. Higher-order terms
also ignored here, since the magnitude of a third-order term is
approximately $1\times 10^{-22}$, much smaller than that of a second-order
term, $1\times 10^{-17}$.

In works by Ashby and colleagues, the terms $\Phi
_{e}/c^{2}-v^{2}/2c^{2}=\Phi _{0}/c^{2}=-6.96927\times 10^{-10}$ were
estimated with the clock fixed to the geoid, without including the tidal
potential of the Moon and Sun \cite%
{ashby2024relativistic,ashby2003relativity}. The IAU agrees with this
definition to within the accuracy needed for the GPS in the terrestrial time
(TT) scale. It is noted that the normalized spherical harmonic coefficients $%
\bar{C}_{l,m}$ and $\bar{S}_{l,m}$ in the Earth's gravitational potential (%
\ref{pe}) tend to smaller values as the degree increases due to the
characteristics of the Earth's gravitational field and the properties of the
Legendre function. Therefore, the higher-degree terms of Earth's
gravitational potential will become smaller, e.g., the term of degree $l=24$
and order $m=1$ giving $\Phi _{e}|_{l=24,m=1}/c^{2}=-1.09983\times 10^{-17}$%
, where $\lambda =0^{\circ }$ and $\phi =86^{\circ }$ are taken. The term
for the Moon's tidal potential can be estimated as $\Phi
_{tm}/c^{2}=-3.89298\times 10^{-17}$, when we set $\psi =0^{\circ }$ in Eq. (%
\ref{tm}). Similarly, the solar tidal potential, another important source of
tidal potential, can be estimated as $\Phi _{ts}/c^{2}=-1.7981\times 10^{-17}
$ when $\psi _{s}=0^{\circ }$ in Eq. (\ref{ts}).

Now, clock measurements have made tremendous progress, thus requiring new
methods for estimating the lunar tidal potential, solar tidal potential, and
higher-degree terms of Earth's gravitational potential. Although the
individual contributions of higher-degree terms of Earth's gravity potential
are only $1\times 10^{-17}$, their cumulative effects can be significant
even at the $1\times 10^{-15}$ level \cite{slava}. This is particularly
important for the ACES (Atomic Clock Ensemble in Space) mission on the
International Space Station, which will operate a clock with accuracy of $%
1\times 10^{-16}$ in low Earth orbit at an altitude of 400 km. Since the
Earth's gravitational potential has been analyzed before, it will not be
further discussed here. In addition, the Earth's orbital period around the
Sun is much longer than the Moon's orbital period around the Earth, and the
Moon's tidal potential on the Earth's surface is several times that of the
Sun. Therefore, we will not discuss the influence of the solar tidal
potential and will focus only on investigating the lunar tidal potential. Of
course, it is not difficult to extend our work to include the solar tidal
potential. Further, studying the lunar tides is also critical for
understanding the lunar gravitational field, since the lunar landing mission
has been brought back to the agenda of current global technological
activities.

\section{Lunar tidal frequency shift between two clocks}

Consideration of the lunar tidal fractional frequency shift between two
clocks on the Earth's surface is important for clock synchronization. Using
the proper time of the geocentric Fermi frame obtained in the previous
section, we can now consider the tidal fractional frequency shift between
the proper times of two clocks on the ground. Therefore, we need to develop
a new comparison method in the local frame of reference for the Earth. In
the geocentric Fermi frame, the proper time of two clocks A and B on the
ground can be obtained according to Eq. (\ref{pc2}) as follows:
\begin{align}
cd\tau _{A}& =\left( 1+\frac{\Phi _{eA}}{c^{2}}+\frac{\Phi _{tmA}}{c^{2}}+%
\frac{\Phi _{tsA}}{c^{2}}-\frac{v_{A}^{2}}{2c^{2}}\right) dx^{0},
\label{pcA} \\
cd\tau _{B}& =\left( 1+\frac{\Phi _{eB}}{c^{2}}+\frac{\Phi _{tmB}}{c^{2}}+%
\frac{\Phi _{tsB}}{c^{2}}-\frac{v_{B}^{2}}{2c^{2}}\right) dx^{0},
\label{pcB}
\end{align}%
where $v_{A}$ and $v_{B}$ are the speeds of clock A and clock B,
respectively, which are caused by the rotation of the Earth. Due to the
difference in geographical latitude and longitude between clocks A and B,
the gravitational potential $\Phi _{e}$, lunar tidal potential $\Phi _{tm}$,
and solar tidal potential $\Phi _{ts}$ differ between the two clocks.

According to Eqs. (\ref{pcA}) and (\ref{pcB}), we can obtain the fractional
frequency shift between A and B as
\begin{equation}
\frac{\triangle f}{f_{B}}=\frac{f_{A}-f_{B}}{f_{B}}=\frac{d\tau _{B}}{d\tau
_{A}}-1=\frac{1+P_{B}}{1+P_{A}}-1,  \label{ff}
\end{equation}%
where $f_{A}=1/d\tau _{A}$, $f_{B}=1/d\tau _{B}$, $P_{A}=\frac{1}{c^{2}}%
\left( \Phi _{eA}+\Phi _{tmA}+\Phi _{tsA}-v_{A}^{2}/2\right) $ and $P_{B}=%
\frac{1}{c^{2}}\left( \Phi _{eB}+\Phi _{tmB}+\Phi _{tsB}-v_{B}^{2}/2\right) $%
. We can expand Eq. (\ref{ff}) to obtain a new form of the fractional
frequency shift between A and B,
\begin{equation}
\frac{\triangle f}{f_{B}}=\left[ 1+P_{B}-\left( 1+P_{B}\right) P_{A}+\left(
1+P_{B}\right) P_{A}^{2}+\cdots \right] -1=P_{B}-P_{A}+O\left( c^{-4}\right)
.
\end{equation}

Moreover the specific form of fractional frequency shift between A and B
with $P_A$ and $P_B$ is
\begin{equation}
\frac{\triangle f}{f_{B}}=\frac{1}{c^{2}}\left( \Phi _{eB}-\Phi _{eA}+\Phi
_{tmB}-\Phi _{tmA}+\Phi _{tsB}-\Phi _{tsA}+\frac{v_{A}^{2}}{2}-\frac{%
v_{B}^{2}}{2}\right) .  \label{ffAB}
\end{equation}

Then, we will analyze the influence of tidal potential on the two clocks on
Earth's surface in the following. Using Eqs. (\ref{tm}) and (\ref{ffAB}),
the fractional frequency shift of the lunar tides between A and B is
\begin{equation}
\frac{\triangle f}{f_{B}}|_{tmAB}=\frac{1}{c^{2}}\left[ \frac{GM_{m}r_{A}^{2}%
}{r_{me}^{3}}\frac{1}{2}\left( 3\cos ^{2}\psi _{tmA}-1\right) -\frac{%
GM_{m}r_{B}^{2}}{r_{me}^{3}}\frac{1}{2}\left( 3\cos ^{2}\psi _{tmB}-1\right) %
\right] ,  \label{fftmAB}
\end{equation}%
where $r_{A}$ and $r_{B}$ are the distances from clock A and clock B to the
center of the Earth, $r_{me}$ is the distance from the Earth to the Moon, $%
\psi _{tmA}$ is the angle between position vector $\mathbf{r}_{A}$ and
position vector $\mathbf{r}_{me}$, and $\psi _{tmB}$ is the angle between
position vector $\mathbf{r}_{B}$ and position vector $\mathbf{r}_{me}$.
Here, the angles $\psi _{tmA}$ and $\psi _{tmB}$ can be computed from the
position vectors $\mathbf{r}_{A}$, $\mathbf{r}_{B}$, and $\mathbf{r}_{me}$.
The position coordinates of clock A and clock B on the Earth's surface and
the Moon can be set as $(x_{A},y_{A},z_{A})=(r_{A}\cos \phi _{A}\cos \lambda
_{A},r_{A}\cos \phi _{A}\sin \lambda _{A},r_{A}\sin \phi _{A})$, $%
(x_{B},y_{B},z_{B})=(r_{B}\cos \phi _{B}\cos \lambda _{B},r_{B}\cos \phi
_{B}\sin \lambda _{B},r_{B}\sin \phi _{B})$ and $(x_{m},y_{m},z_{m})=(r_{me}%
\cos \phi _{m}\cos \lambda _{m},r_{me}\cos \phi _{m}\sin \lambda
_{m},r_{me}\sin \phi _{m})$, where $\phi $ is the latitude of the position
vectors and $\lambda $ is the longitude of the position vectors. Here, the
coordinates $(x_{A},y_{A},z_{A})$ of clock A and $(x_{B},y_{B},z_{B})$ of
clock B are spherical coordinates on the Earth's surface. The coordinates $%
(x_{m},y_{m},z_{m})$ of the Moon's position are spherical coordinates on a
sphere centered at the Earth's center and with radius $r_{me}$. Note that $%
r_{A,B}$ are much smaller than $r_{me}$, meaning the Moon is on a sphere
larger than the Earth's surface. In order to conveniently consider the tidal
fractional frequency shift between clocks A and B on the ground, the
longitude and latitude of the Moon relative to the Earth are used for
analysis \cite{murray1999solar}. We can then obtain the angles $\psi _{tmA}$
and $\psi _{tmB}$ as follows:
\begin{align}
\cos \psi _{tmA}& =\frac{x_{m}x_{A}+y_{m}y_{A}+z_{m}z_{A}}{r_{A}r_{me}}=\sin
\phi _{A}\sin \phi _{m}+\cos \phi _{A}\cos \phi _{m}\cos \left( \lambda
_{A}-\lambda _{m}\right) ,  \label{pA} \\
\cos \psi _{tmB}& =\frac{x_{m}x_{B}+y_{m}y_{B}+z_{m}z_{B}}{r_{B}r_{me}}=\sin
\phi _{B}\sin \phi _{m}+\cos \phi _{B}\cos \phi _{m}\cos \left( \lambda
_{B}-\lambda _{m}\right) .  \label{pB}
\end{align}

Through Eqs. (\ref{fftmAB}), (\ref{pA}), and (\ref{pB}), we can
obtain the specific expression of the lunar tidal fractional frequency shift
between clocks A and B as
\begin{equation}
\begin{split}
\frac{\triangle f}{f_{B}}|_{tmAB}& =\frac{3}{4}\frac{GM_{m}r_{A}^{2}}{%
c^{2}r_{me}^{3}}\left[ \cos ^{2}\phi _{A}\cos ^{2}\phi _{m}\cos 2\left(
\lambda _{A}-\lambda _{m}\right) +\sin 2\phi _{A}\sin 2\phi _{m}\cos \left(
\lambda _{A}-\lambda _{m}\right) \right.  \\
& \left. +3\left( \sin ^{2}\phi _{A}-\frac{1}{3}\right) \left( \sin ^{2}\phi
_{m}-\frac{1}{3}\right) \right] -\frac{3}{4}\frac{GM_{m}r_{B}^{2}}{%
c^{2}r_{me}^{3}}\left[ \cos ^{2}\phi _{B}\cos ^{2}\phi _{m}\cos 2\left(
\lambda _{B}-\lambda _{m}\right) \right.  \\
& \left. +\sin 2\phi _{B}\sin 2\phi _{m}\cos \left( \lambda _{B}-\lambda
_{m}\right) +3\left( \sin ^{2}\phi _{B}-\frac{1}{3}\right) \left( \sin
^{2}\phi _{m}-\frac{1}{3}\right) \right]. \label{ltm}
\end{split}%
\end{equation}

For convenience of comparison, we put clock A at $0^{\circ }$ longitude or $%
0^{\circ }$ latitude to compare with clock B. Therefore, we consider that
the longitude $\lambda _{A,B}$ range for the clock on the ground is from $%
0^{\circ }$ to $180^{\circ }$, and the latitude $\phi _{A,B}$ range is from $%
0^{\circ }$ to $90^{\circ }$. The Moon's orbital plane around Earth lies
closer to the ecliptic than to Earth's celestial equator. Because the Moon's
orbit is tilted by approximately $5^{\circ }$ relative to the ecliptic plane
of Earth's orbit around the Sun \cite{simon1994numerical}, the Moon's
orbital inclination varies in space, increasing or decreasing relative to
Earth's axial tilt of $23.5^{\circ }$ \cite{standish1982orientation,
park2021jpl}. The Moon's maximum inclination varies from $+28.5^{\circ }$ to
$-28.5^{\circ }$, for a total range of $57^{\circ }$. This range occurs over
a short period of approximately 2 weeks. Therefore, the latitude $\phi _{m}$
range of the Moon is from $-28.5^{\circ }$ to $28.5^{\circ }$, and the
longitude $\lambda _{m}$ range is from $0^{\circ }$ to $360^{\circ }$.

\begin{figure}[tbp]
\centering
{\includegraphics[width=0.76\textwidth]{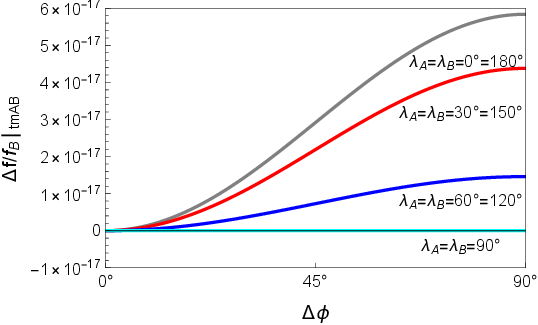}}
\caption{The Moon tidal fractional frequency shift between clocks A and B,
which have the same longitude but different latitudes. The longitude values
are located at $\protect\lambda _{A}=\protect\lambda _{B}=0^{\circ
},30^{\circ },60^{\circ },90^{\circ },120^{\circ },150^{\circ },180^{\circ }$
. Other parameters are in SI units: $M_{m}=7.346\times 10^{22}$, $%
G=6.674\times 10^{-11}$, $c=3\times 10^{8}$, $\protect\lambda _{m}=0^{\circ
} $, $\protect\phi _{m}=0^{\circ }$, $r_{A}=r_{B}=6.371\times 10^{6}$, and $%
r_{me}=3.843\times 10^{8}$.}
\label{fig1}
\end{figure}

Vessot et al. analyzed the fractional frequency shift between a ground-based
clock and a satellite clock caused by the clock's motion and the Earth's
gravitational potential in a $c^{-2}$ post-Newton geocentric framework \cite%
{vessot1980test}. Blanchet et al. further expanded the research to the $%
c^{-3}$ order \cite{blanchet2001relativistic}, but neither of them took into
account the influence of the tidal potential of external massive celestial
bodies on the clock. Zhang et al. studied the fractional frequency shift
between clocks on the lunar surface and those on the Earth's surface \cite%
{zhang2025frequency}. They considered that the factors causing the
fractional frequency shifts between the clocks include the tidal potential
caused by massive celestial bodies in the Solar System. Qin et al. \cite%
{qin2020tidal} considered the fractional frequency shift between two clocks
at the same longitude but different latitudes on Earth, caused by the tidal
potential of the Sun and Moon. They further provided a schematic diagram of
the tidal clock effect, which is dependent on the clock's latitude, for
comparison purposes in a global clock network, but the clock's longitude was
not included.

\begin{figure}[ptb]
\centering
{\includegraphics[width=0.76\textwidth]{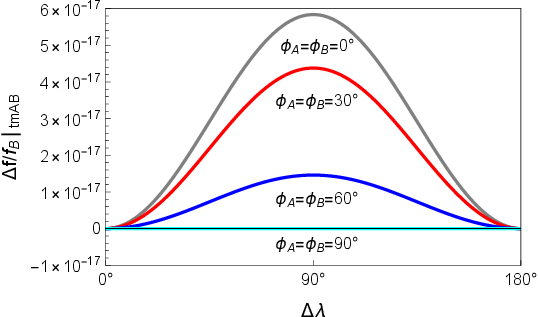}}
\caption{The Moon tidal fractional frequency shift between clocks A and B,
which have the same latitude but different longitudes. The four different
latitude values are located at $\protect\phi_A=\protect\phi_B=0^\circ,
30^\circ, 60^\circ, 90^\circ$. Other parameters are kept consistent with
those in Fig. \protect\ref{fig1}}
\label{fig2}
\end{figure}

According to our calculations, we can include the lunar longitude $\lambda
_{m}$, the lunar latitude $\phi _{m}$, the longitude of clocks $\lambda _{A}$%
, and the latitude of clocks $\phi _{A}$ in the angle $\cos \psi _{tmA}=\sin
\phi _{A}\sin \phi _{m}+\cos \phi _{A}\cos \phi _{m}\cos \left( \lambda
_{A}-\lambda _{m}\right) $ between position vectors $\mathbf{r}_{A}$ and $%
\mathbf{r}_{me}$. This approach allows us to analyze the change in lunar
tidal fractional frequency shift caused by the longitude difference between
two ground clocks. We can also analyze the change in lunar tidal fractional
frequency shift caused by changes in lunar longitude and latitude. Although
the reference frames we consider are different, when moving a clock on the
surface of the Earth to the surface of the Moon, we can also estimate the
fractional frequency shift between the clock on the surface of the Moon and
the clock on the surface of the Earth as in Vessot et al. \cite%
{vessot1980test}, Blanchet et al. \cite{blanchet2001relativistic}, and Zhang
et al. \cite{zhang2025frequency}.

In Fig. \ref{fig1}, we consider two ground clocks located at the same
longitude but different latitudes, with clock A being located at $0^{\circ }$
latitude. We obtain results consistent with those of Qin et al. \cite%
{qin2020tidal,qin2023tidal}. When the two clocks are at the same longitude,
the lunar tidal fractional frequency shift increases with the latitude
difference $\triangle \phi $ between them, reaching a maximum of
approximately $6\times 10^{-17}$ when the clocks are located at the equator
and the North Pole, respectively.

\begin{figure}[ptb]
\centering
\begin{subfigure}[h]{0.48\textwidth}
		{\includegraphics[width=\textwidth]{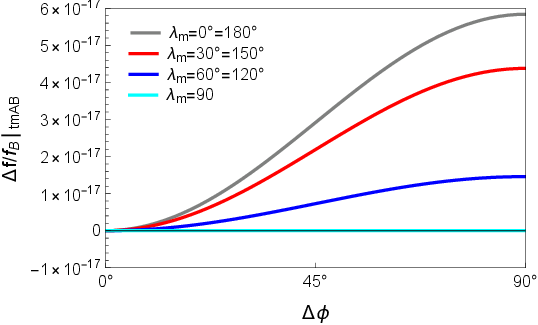}}
		\caption{}
		\label{3a}
	\end{subfigure}
\begin{subfigure}[h]{0.49\textwidth}
		{\includegraphics[width=\textwidth]{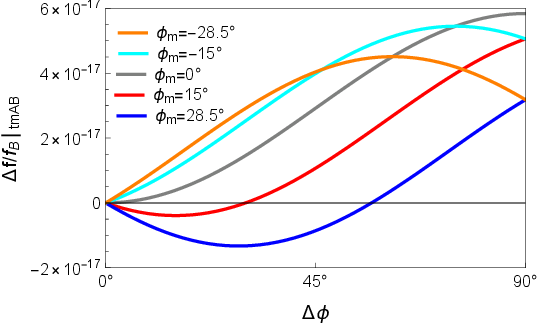}}
		\caption{}
		\label{3b}
	\end{subfigure}
\caption{The Moon tidal fractional frequency shift between clocks A and B
with same longitude $\protect\lambda_A=\protect\lambda_B=0^\circ$ but
different latitudes, where (a) choose the same Moon latitude $\protect\phi %
_m=0^\circ$, but the different Moon longitudes $\protect\lambda %
_m=0^\circ,30^\circ, 60^\circ,90^\circ ,120^\circ,150^\circ,180^\circ$ and
(b) choose the same Moon longitude $\protect\lambda_m=0^\circ$, but the
different Moon latitudes $\protect\phi_m=0^\circ,-15^\circ,-28.5^\circ,15^
\circ,28.5^\circ$. Other parameters are in SI units: $M_m=7.346\times
10^{22} $, $G=6.674\times10^{-11}$, $c=3\times 10^8$, $r_A=r_B=6.371 \times
10^6$, and $r_{me}=3.843\times 10^8$.}
\label{fig3}
\end{figure}

An important distinction is that the lunar tidal fractional frequency shift
between the two clocks increases when the longitude decreases from $%
90^{\circ }$ to $0^{\circ }$ or increases towards $180^{\circ }$ when the
latitude difference $\triangle \phi $ between the two clocks is the same.
Another point to note is that since we set $\lambda _{m}=0^{\circ }$ and $%
\phi _{m}=0^{\circ }$, when the latitude of the two clocks is at $\lambda
_{A}=\lambda _{B}=90^{\circ }$, $\cos \psi _{tmA}$ and $\cos \psi _{tmB}$
will be equal to $0$. At this time, there is no lunar tidal fractional
frequency shift between the two clocks. This can be obtained by simplifying
Eq. (\ref{ltm}) into
\begin{equation}
\frac{\triangle f}{f_{B}}|_{fig.1}=\cos \left( 2\lambda _{A/B}\right) \left[
2.91\times 10^{-17}-2.91\times 10^{-17}\cos ^{2}\left( \triangle \phi
\right) \right] +2.91\times 10^{-17}\sin ^{2}\left( \triangle \phi \right) .
\label{eq.2}
\end{equation}%
Thus, it is easier to see the result of the function $\cos \left( 2\lambda
_{A/B}\right) $ modulation on the amplitude term of the lunar tidal
fractional frequency shift.

\begin{figure}[ptb]
\centering
\begin{subfigure}[h]{0.49\textwidth}
		{\includegraphics[width=\textwidth]{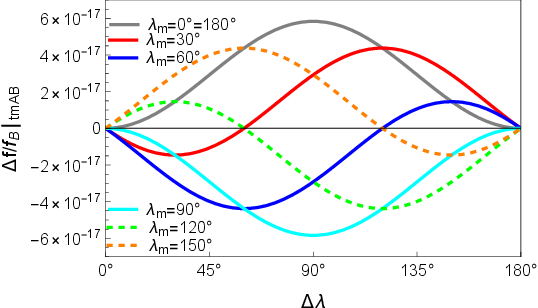}}
		\caption{}
		\label{4a}
	\end{subfigure}
\begin{subfigure}[h]{0.49\textwidth}
		{\includegraphics[width=\textwidth]{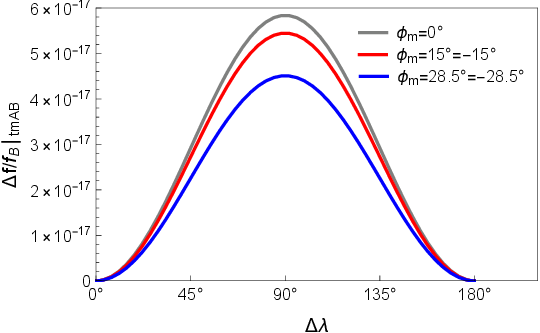}}
		\caption{}
		\label{4b}
	\end{subfigure}
\caption{The Moon tidal fractional frequency shift between clocks A and B
with same latitude $\protect\phi_A=\protect\phi_B=0^\circ$ but different
longitudes, where (a) choose the same Moon latitude $\protect\phi_m=0^\circ$
, but the different Moon longitudes $\protect\lambda_m=0^\circ,30^\circ,
60^\circ,90^\circ ,120^\circ,150^\circ,180^\circ$ and (b) choose the same
Moon longitude $\protect\lambda_m=0^\circ$, but the different Moon latitudes
$\protect\phi_m=0^\circ,-15^\circ,-28.5^\circ,15^\circ,28.5^\circ$. Other
parameters are kept consistent with those in Fig.3.}
\label{fig4}
\end{figure}

We found that the lunar tidal fractional frequency shift of two clocks
located at the same latitude but different longitudes increases with an
increase in the longitude difference from $0^{\circ }$ to $90^{\circ }$ or a
decrease from $180^{\circ }$ to $90^{\circ }$. Furthermore, when the
longitude difference $\triangle \lambda $ between the two clocks is the
same, the lunar tidal fractional frequency shift increases as the latitude
of the two clocks decreases. It can be seen that the function graph of the
lunar tidal fraction frequency shift in the figure exactly matches the
equation%
\begin{equation}
\frac{\triangle f}{f_{B}}|_{fig.2}=\cos ^{2}\left( \phi _{A/B}\right) \left[
2.91\times 10^{-17}-2.91\times 10^{-17}\cos \left( 2\triangle \lambda
\right) \right] ,  \label{eq.3}
\end{equation}%
which is obtained by simplifying Eq. (\ref{ltm}). In Fig. \ref{fig2}, we can
see that when $\phi _{A}=\phi _{B}=90^{\circ }$, $\cos \psi _{tmA}$ and $%
\cos \psi _{tmB}$ are equal to $0$, the lunar tidal fractional frequency
shift between the two clocks is zero. Another, more straightforward
explanation is that when the two clocks coincide and are located at the
North Pole, there is no lunar tidal fractional frequency shift.

Figures \ref{fig1} and \ref{fig2} consider the changes in the lunar tidal
fractional frequency shift caused by the Moon at a fixed position $\lambda
_{m}=0^{\circ },\phi _{m}=0^{\circ }$, and two clocks at different locations
on the Earth's surface. Figure \ref{fig3} shows how the lunar tidal
fractional frequency shift changes when the Moon's longitude and latitude
are varied for two clocks at the same longitude but different latitudes. In
Fig. \ref{3a}, decreasing the lunar longitude from $90^{\circ }$ to $%
0^{\circ }$ or increasing it from $90^{\circ }$ to $180^{\circ }$ increases
the lunar tidal fractional frequency shift for the two clocks with the same
latitude difference $\triangle \phi $. The reason for this change can be
seen from
\begin{equation}
\frac{\triangle f}{f_{B}}|_{fig.3a}=\cos \left( 2\lambda _{m}\right) \left[
2.91\times 10^{-17}-2.91\times 10^{-17}\cos ^{2}\left( \triangle \phi
\right) \right] +2.91\times 10^{-17}\sin ^{2}\left( \triangle \phi \right)
\label{eq.4a}
\end{equation}%
which is obtained from Eq. (\ref{ltm}). In this case, changing the longitude
of the Moon only changes the amplitude term $\cos \left( 2\lambda
_{m}\right) $ of the lunar tidal fraction frequency shift. Here, we find
that the lunar longitude changes with a period of $180^{\circ }$ between the
two clocks of the lunar tide, so hereafter we only consider the range of
lunar longitude changes from $0^{\circ }$ to $180^{\circ }$. In Fig. \ref{3b}%
, the change of the Moon's latitude leads to the change of the magnitude and
phase of the lunar tidal fractional frequency shift for the two clocks with
the same latitude difference $\triangle \phi $. A more intuitive reason can
be seen from
\begin{equation}
\frac{\triangle f}{f_{B}}|_{fig.3b}=2.91\times 10^{-17}\cos \left( 2\phi
_{m}\right) -2.91\times 10^{-17}\cos \left( 2\phi _{m}-2\triangle \phi
\right)   \label{eq.4b}
\end{equation}%
which is obtained from Eq. (\ref{ltm}). In this case, changing the latitude
of the Moon will simultaneously change the amplitude and phase of the lunar
tidal fractional shift function.

In Fig. \ref{fig4}, we consider the changes in the lunar tidal fractional
frequency shift caused by variations in the Moon's longitude and latitude
when the two clocks are located at the same latitude but different
longitudes. In Fig. \ref{4a}, the Moon's latitude changes with a period of $%
180^{\circ }$, resulting in both the magnitude and phase of the lunar tidal
fractional frequency shift of the two clocks with the same longitude
difference $\triangle \lambda $. In Fig. \ref{4b}, changing the Moon's
latitude only changes the magnitude of the lunar tidal fractional frequency
shift of the two clocks with the same longitude difference $\triangle
\lambda $. Similarly, we can simplify Eq. (\ref{ltm}) into%
\begin{equation}
\frac{\triangle f}{f_{B}}|_{fig.4a}=2.91\times 10^{-17}\cos \left( 2\lambda
_{m}\right) -2.91\times 10^{-17}\cos \left( 2\triangle \lambda -2\lambda
_{m}\right) ,  \label{eq.5a}
\end{equation}

\begin{equation}
\frac{\triangle f}{f_{B}}|_{fig.4b}=\cos ^{2}\left( \phi _{m}\right) \left[
2.91\times 10^{-17}-2.91\times 10^{-17}\cos \left( 2\triangle \lambda
\right) \right] .  \label{eq.5b}
\end{equation}%
They show that when the latitudes of the two clocks on the ground are the
same but their longitudes are different, the lunar longitude will appear in
the amplitude and phase of the lunar tidal fractional frequency shift
function, thus obtaining the result in Fig. \ref{4a}. The lunar latitude
will only appear in the amplitude of the lunar tidal fractional frequency
shift function, thus obtaining the result in Fig. \ref{4b}.

\section{Conclusions}

In this paper, we studied the fractional frequency shift due to lunar tides
between clocks at two different locations on Earth. When the position of the
Moon is fixed, for clocks on Earth at the same longitude, increasing the
latitude difference increases the fractional frequency shift; for clocks on
Earth at the same latitude difference $\triangle \phi $, the fractional
frequency shift increases as the longitude decreases from $90^{\circ }$ to $%
0^{\circ }$ or increases from $90^{\circ }$ to $180^{\circ }$. Further, we
find that for clocks on Earth at the same latitude, the fractional frequency
shift increases as the longitude difference increases between $0^{\circ }$
and $90^{\circ }$ or as the longitude difference decreases between $%
90^{\circ }$ and $180^{\circ }$; and for clocks at the same longitude
difference $\triangle \lambda $, the fractional frequency shift also
increases as the latitude of the clock decreases.

In addition to the position of the two clocks on the ground, the changes in
longitude and latitude caused by the movement of the Moon also affect the
fractional frequency shift between the two clocks. It is found that for two
clocks at the same latitude but with a longitude difference $\triangle
\lambda $, the change of the Moon's longitude will lead to a change of the
phase and magnitude of the lunar tidal fractional frequency shift for both
clocks, while a change of the lunar latitude only affects the magnitude of
the lunar tidal fractional frequency shift. Interestingly, for two clocks
located at the same longitude but at different latitudes, changing the lunar
longitude only affects the magnitude of the lunar tidal fractional frequency
shift, while changing the lunar latitude alters both the phase and magnitude
of the lunar tidal fractional frequency shift for both clocks. The position
dependence of the lunar tidal frequency shift between two ground clocks
complements the lunar tidal effect. These effects are of great significance
for the calibration and synchronization of ground clocks.

\section*{Acknowledgments}

This work is supported by National Natural Science Foundation of China
(NSFC) with Grant No. 12375057, and the Fundamental Research Funds for the
Central Universities, China University of Geosciences (Wuhan).

\end{document}